\documentclass[amsmath, amsfonts, superscriptaddress, showpacs, prl]{revtex4}
\usepackage{graphicx}
\usepackage{epsfig}
\usepackage{bm}
\usepackage{dcolumn}
\usepackage{amsmath}
\usepackage{amssymb}
\usepackage{wrapfig}

\newcommand{\arccosh}{\mathop{\rm arccosh}\nolimits}

\begin{document}

\title{Enhancement of the London penetration depth in pnictides
at the onset of SDW order under superconducting dome}

\author{A.~Levchenko}
\affiliation{Department of Physics and Astronomy, Michigan State
University, East Lansing, Michigan 48824, USA}

\author{M.~G.~Vavilov}
\affiliation{Department of Physics, University of Wisconsin,
Madison, Wisconsin 53706, USA}

\author{M.~Khodas}
\affiliation{Department of Physics and Astronomy,
University of Iowa, Iowa City, Iowa 52242, USA}

\author{A.~V.~Chubukov}
\affiliation{Department of Physics, University of Wisconsin,
Madison, Wisconsin 53706, USA}

\begin{abstract}
Recent measurements of the doping dependence of the London
penetration depth $\lambda(x)$ at low $T$ in clean samples of
isovalent BaFe$_2$(As$_{1-x}$P$_x$)$_2$ at $T \ll T_c$ [Hashimoto
\textit{et al.}, Science \textbf{336}, 1554 (2012)] revealed a peak
in $\lambda (x)$  near optimal doping $x=0.3$. The observation of
the peak at $T\ll T_c$, points to the existence of the quantum
critical point (QCP) beneath the superconducting dome. We associate
such a QCP with the onset of a spin-density-wave order and show that
the renormalization of $\lambda (x)$ by critical magnetic
fluctuations gives rise to the observed feature. We argue that the
case of pnictides is conceptually different from a one-component
Galilean invariant Fermi liquid, for which correlation effects do
not cause the renormalization of the London penetration depth at
$T=0$.
\end{abstract}

\date{April 25, 2013}


\maketitle

\textit{Introduction}.-- The properties of iron-based
superconductors (FeSCs) have been at the forefront of research
activities in the correlated electron community over the last few
years~\cite{Exp-1,Exp-2,Exp-3,Exp-4}. These materials have multiple
Fermi pockets with electronlike and holelike dispersion of carriers.
It is well established that superconductivity in FeSCs emerges in
close proximity to  a spin-density-wave (SDW) order, and the
superconducting (SC) critical temperature $T_c$ has dome-shaped
dependence on doping, with $T_c$ maximum near the onset of SDW
order~\cite{PD-Exp-1,PD-Exp-2,PD-Exp-3,PD-Exp-4}.

Several groups~\cite{rev} put forward the scenario that
superconductivity in FeSCs  has $s^{+-}$ symmetry and emerges
because SDW fluctuations increase interpocket interaction, which is
attractive for $s^{+-}$ gap symmetry, to a level when it overcomes
intrapocket repulsion. Likewise, SC fluctuations tend to increase
the tendency towards SDW.

Once the system develops long-range order, the situation changes
because SDW and SC orders compete, and the order which sets first
tends to block the development of the other. According to theory,
such competition may give rise to a homogeneous coexistence of SDW
and SC orders in some range of
dopings~\cite{Fernandes-PRB10,Vorontsov-SUST,Vorontsov-PRB10}. A
homogeneous coexistence of SDW and SC orders has been detected in
122 materials -- electron-doped
Ba(Fe$_{1-x}$Co$_x$)$_2$As$_2$~\cite{LP-Exp-1,LP-Exp-2,LP-Exp-3,PD-Exp-3,PD-Exp-4,BM-Exp-1,BM-Exp-2,BM-Exp-3}
and hole-doped
Ba$_{1-x}$K$_x$Fe$_2$As$_2$~\cite{BM-Exp-4,BM-Exp-5,BM-Exp-6}. On
the other hand, for EuFe$_{2-x}$Co$_x$As$_2$ M\"{o}ssbauer
spectroscopy measurements~\cite{Karpinski-PRB11} were interpreted in
favor of phase separation, when SC has a filamentary character and
is concentrated in nonmagnetic regions. In the third class of 122
materials -- an isovalent BaFe$_2$(As$_{1-x}$P$_x$)$_2$, the
coexistence between SDW and SC order has not yet been probed
experimentally, but the odds are that the two orders do coexist
because the phase diagram of BaFe$_2$(As$_{1-x}$P$_x$)$_2$  is quite
similar to that for Ba(Fe$_{1-x}$Co$_x$)$_2$As$_2$~\cite{Matsuda-1}.

\begin{wrapfigure}{r}{0.45\textwidth}
\begin{center}
\vspace{-1cm}
  \includegraphics[width=0.45\textwidth]{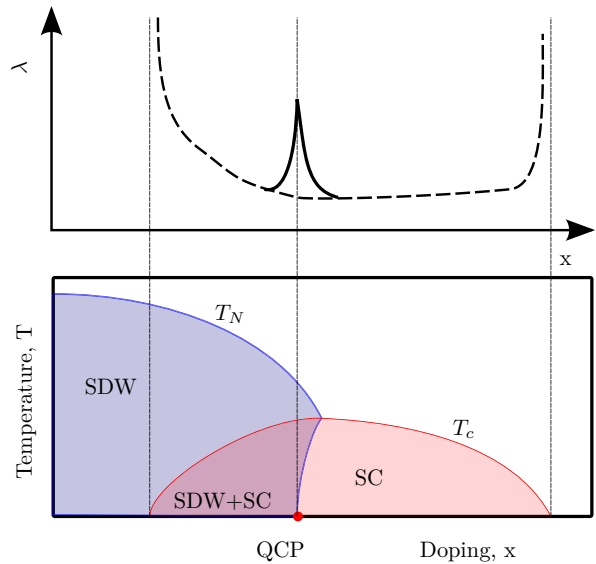}
\end{center}\vspace{-.45cm}
\caption{Lower panel: a theoretical  phase diagram of 122-type
iron-based superconductors in temperature \textit{vs} doping planes.
Critical temperatures $T_N$ and $T_c$ indicate transitions into pure
SDW and SC phases, respectively. A QCP lies beneath the SC dome and
separates pure SC and coexistence SC+SDW phases. Reentrant behavior
of $T_N$ under the SC dome has been detected in Co-doped
 122 materials~\cite{new} but well may be
 nonuniversal~\cite{comm,Fernandes-PRB10,Vorontsov-SUST,Vorontsov-PRB10}.
Upper panel: the theoretical behavior of the penetration depth
$\lambda$ at $T=0$. In the mean-field approximation (dashed line),
$\lambda$ diverges at the edges of the superconducting dome, flat
inside pure SC phase and increases monotonically as the system moves
towards the pure SDW phase. Beyond the mean field, magnetic
fluctuations give rise to a peak in $\lambda$ at the onset of the
SDW order (solid line). The peak in $\lambda$ has been observed in
Ref. ~\cite{Matsuda-2}.} \label{Fig-PD}
\end{wrapfigure}

The coexistence implies that the SDW transition line extends into
the superconducting phase. If this line reaches $T=0$, the system
develops a magnetic quantum-critical point (QCP) beneath the
superconducting dome~\cite{comm}, see Fig.~\ref{Fig-PD}. A magnetic
QCP without superconductivity has been analyzed in great
detail~\cite{qcp,qcp_1}, and it is known that quantum fluctuations
near this point give rise to non-Fermi liquid (NFL) behavior and to
singularities in various electronic characteristics. An SDW
instability inside the $d$-wave SC state has been analyzed
in~\cite{subir_1} and was shown to give rise to NFL behavior of
nodal fermions.

The observation of coexistence brings about the new issue of whether
there are electronic singularities at a magnetic QCP which develops
in the presence of an $s^{+-}$ SC order. Of particular interest are
the singularities in quantities such as the penetration depth
$\lambda (x)$, which measures electronic response averaged over the
whole Fermi surface (FS). Early experiments~\cite{ruslan_11} on
Ba(Fe$_{1-x}$Co$_x$)$_2$As$_2$ found no special features in
$\lambda(x)$ at the onset of SDW order, but recent measurements in
BaFe$_2$(As$_{1-x}$P$_x$)$_2$ (Ref.~\onlinecite{Matsuda-2}) found a
peak in $\lambda (x)$ at the smallest $T\ll T_c$ at around optimal
doping (see inset of Fig.~\ref{Fig-PD}). The authors of
Ref.~\cite{Matsuda-2} speculated that the peak likely indicates that
there is a QCP beneath a SC dome and argued that the peak in
$\lambda (x)$ is a generic feature of 122 Fe pnictides, but it is
more difficult to detect it in Ba(Fe$_{1-x}$Co$_x$)As$_2$ because of
the greater degree of electronic disorder caused by Co doping.
Another potential reason why the peak has been observed only in
BaFe$_2$(As$_{1-x}$P$_x$)$_2$ is that this material possesses gap
nodes~\cite{Matsuda_nodes}, which generally lead to stronger effects
due to quantum fluctuations.

In this Letter, we analyze the behavior of $\lambda (x)$ under the
assumption that the QCP is associated with the development of SDW
order beneath a superconducting dome. A preemptive nematic order may
also play a role~\cite{Fernandes-PRB12}, but we will not dwell on
that.

\textit{London penetration depth near QCP}. -- In general, the peak
in $\lambda (x)$ at a SDW QCP can emerge for one of three reasons:
(i) a nonmonotonic behavior of $\lambda$ near a QCP already within
the mean-field theory (like the peak in the specific heat jump at
$T_c$ at the onset of coexistence with SDW~\cite{Vavilov-PRB11}),
(ii) critical fluctuations at the onset of SDW, not specific to the
form of the gap, and (iii) critical fluctuations specific to the
presence of the gap nodes. Besides, $\lambda$ can either diverge at
a QCP, or get enhanced but stay finite. It was found
recently~\cite{Dushko} that, at the mean field level, the variation
of $\lambda$ is smooth and cannot explain sharp features observed in
Ref.~\cite{Matsuda-2}. Here, we investigate the effects of critical
magnetic fluctuations. We find that fluctuations associated with SDW
QCP beneath a SC dome give rise to the enhancement of the effective
mass $m^*$. The mass does not diverge because SC order cuts infrared
singularities, but, nevertheless, $m^*/m$ at the QCP is noticeably
enhanced. We argue that the enhancement of $m^*$ gives rise to a
sharp peak in $\lambda(x)$ at the onset of coexistence with SDW. We
also find that the presence of the nodes in the gap is not
sufficient to transform a peak into a divergence because  the
dominant contribution to $m^*$ comes from the region away from the
nodes.

London penetration depth in a type-II superconductor with cubic
symmetry is expressed via the zero-momentum component of the
electromagnetic response tensor $Q_{ij}({\bm{k}}) = (\delta_{ij} -
k_ik_j/k^2) Q({\bm{k}})$, which relates vector potential ${\bm{A}}$
and the current density ${\bm{j}}$: $j_i (\bm{k}) = -Q_{ij}
({\bm{k}}) A_j ({\bm{k}})$. The temperature and doping dependent
penetration depth is given by $\lambda^{-2}(T,x)=(4\pi/c)Q (T,x)$,
where  $c$ is the velocity of light. The kernel $Q(T,x) $ is related
to the current-current correlation function in the limit of zero
frequency and vanishing momentum and is expressed via the superfluid
density $n_s (T,x)$ as $ Q = e^2 n_s/mc$, where $m$ and $e$ are the
mass and the charge of an electron. Then $\lambda^2 =mc^2/(4\pi e^2
n_s)$. In the Galilean invariant, a one-component fermionic system
superfluid density at $T=0$ is equal to the total density of
fermions $n(x)$. In this situation, $\lambda (T=0,x)$ does not
depend on Fermi liquid corrections and remains the same as in a
Fermi gas~\cite{Larkin,Leggett}. Diagrammatically, superfluid
density is given by the sum of two bubble diagrams made out of
normal and anomalous Green's functions, and the independence of $n_s
(T=0,x)$ on the electron-electron interaction is the result of the
cancelation between self-energy and vertex corrections to these
diagrams. At $T>0$ the $T-$dependent part of $n_s$ does depend on
Fermi liquid parameters~\cite{Leggett}.

We find, however, that in iron pnictides the situation is different
because these systems have multiple Fermi pockets, and $s^{+-}$
pairing originating from interpocket interaction. The interplay
between self-energy and vertex corrections then depends on the
orientation of Fermi velocities and the values of superconducting
order parameters at different FSs. We find that self-energy and
vertex  corrections generally do not cancel, and the penetration
depth is roughly proportional to $m^*/m$.

We followed earlier works~\cite{rev} and assumed that the most
relevant interaction in Fe pnictides is between hole and electron
pockets, separated by ${\bf Q}=(\pi,\pi)$ in the folded Brillouin
zone, and that the gap has $s^{+-}$ symmetry and changes sign
between electron and hole pockets. We calculated the leading
interaction correction to $\lambda (x)$ in the one-loop
approximation. This perturbative analysis is justified because
renormalized $\lambda (x)$ does not diverge even at a SDW QCP. There
are 16 diagrams with one-loop corrections to current-current
correlators, half of them are self-energy and half are vertex
corrections. We evaluated the diagrams and found that self-energy
and vertex corrections are of the same order, and both decrease the
superfluid density and increase the penetration depth~\cite{SM}. To
be brief, below we analyze how $\lambda (x)$ is affected by
inserting fermionic self-energy into the current correlation
function. A straightforward calculation yields, at one-loop order
\begin{equation}\label{Z}
\lambda^{2}(T=0,x)= \lambda^{2}_{BCS} \left[1+ \beta(x)\right],
\end{equation}
where
\begin{equation}\label{Z_1}
\beta = \Big\langle \sum_j \left[1- Z_j
(\bm{k}_F)\right]\Big\rangle_\phi\!\!\!\! = \left\langle
\lim_{\omega\to
0}\partial_{i\omega_m}\Sigma_j(\bm{k}_F,\omega_m)\right\rangle_\phi.
\end{equation}
Here, $j$ labels Fermi pockets, $\Sigma$ is a diagonal (normal)
self-energy, which generally depends on the location of ${\bm{k}_F}$
on the corresponding Fermi surface, and
$\langle\ldots\rangle_\phi=\int_0^{2\pi}\ldots d\phi/2\pi$. In a
situation where the dependence of $\Sigma_j(\bm{k},\omega_m)$ on
${\bm k}-{\bm k}_F$ can be neglected, the quasiparticle residue is
related to mass renormalization as $ Z_j (\bm{k}_F) = m/m^*_j
({\bm{k}_F})$.

A similar expression for $\lambda$ has been obtained earlier for
heavy-fermion superconductor UBe$_{13}$~\cite{Varma}, which is a
two-component system of conduction $d-$ electrons and localized $f$
electrons, of which only the first carry the current.  It is
tempting to extend the one-loop result \eqref{Z} to $\lambda^{-2}
\propto \sum_j m/m^*_j$, but we caution that  noncancellation of
one-loop self-energy and vertex corrections to the current
correlator does not necessarily imply that vertex corrections can be
simply neglected. An example of more complex behavior beyond
one-loop order has been recently considered in~\cite{max_last}.

\textit{Evaluation of the fermionic self-energy}.-- We consider the
minimal three-band model of two elliptical electron Fermi surfaces
and one circular hole Fermi surface. The basic Hamiltonian includes
the free fermion part $H_0$ and pair fermion interactions in
superconducting $H_\Delta$ and magnetic $H_\sigma$
channels~\cite{SM}. These interactions are described by the local
coupling constants $g_{\rm sc}$ and $g_{\rm sdw}$ respectively. The
phase diagram of the model has been obtained
before~\cite{Vorontsov-PRB10}. We focus on the region where at $T=0$
the system has a long-range SC order and is about to develop an SDW
order. Renormalization of mass on all Fermi surfaces is of the same
order, and for brevity we show the calculations of $m^*/m$ for just
one pocket.

Potentially, singular self-energy comes from the exchange of
near-critical SDW fluctuations.  In the normal state, these
fluctuations are overdamped and are slow compared to electrons. In a
SC state, the dynamical exponent changes from $z=2$ to $z=1$ because
fermions which contribute to bosonic dynamics become massive
particlelike excitations. Such systems have been discussed earlier
in the context of cuprates~\cite{qcp} and we follow the same
approach in deriving the expressions for the self-energy and spin
polarization operator in the SC state in our case.

The one-loop self-energy due to spin-fluctuation exchange is a
convolution of spin-fluctuation and fermionic propagators, both
taken in the superconducting state:
\begin{equation}
\Sigma_j(\bm{k},\omega_n)=3 T\sum_{\Omega_m} \int
\frac{d\bm{q}}{4\pi^2}
L(\bm{q},\Omega_m)\mathcal{G}_j(\bm{k}-\bm{q},\omega_n-\Omega_m)
\end{equation}
where $\omega_m=2\pi T(n+1/2)$ and $\Omega_m=2\pi mT$ are fermionic
and bosonic Matsubara frequencies respectively. The normal and
anomalous components of the Green's function in the SC are
\begin{equation}\label{G}
\mathcal{G}_j(\bm{k},\omega_n)=\frac{-i\omega_n-\xi_j}{\xi^2_j+\omega^2_n+\Delta^2_j},~~
\mathcal{F}_j(\bm{k},\omega_n)=\frac{-\Delta_j}{\xi^2_j+\omega^2_n+\Delta^2_j}~
\end{equation}
where $\xi_j=\xi_j(\bm{k})=\bm{v}_{j,F} (\bm{k}-\bm{k}_F)$,
 and the energy gap $\Delta_j$ is equal to $\Delta_h$ on
the hole Fermi surfaces and $\Delta_e (\phi) =
-\Delta_e(1\pm\alpha\cos2\phi)$ on the two electron Fermi surfaces
(we choose $\Delta_h, \Delta_e >0$). The gaps on electron pockets
have nodes when $\alpha >1$. We emphasize that the SC gap can be
treated as doping independent only in the paramagnetic state. Once
SDW order sets in, the value of the gap
changes~\cite{Fernandes-PRB10,Vorontsov-SUST,Vorontsov-PRB10}.

The spin-fluctuation propagator is given by
\begin{equation}\label{L}
L(\bm{q},\Omega_m)=\frac{1}{g^{-1}_{{\rm
sdw}}+\Pi(\bm{q},\Omega_m)},
\end{equation}
where the polarization operator $\Pi(\bm{q},\Omega_m)$ is (see
\cite{SM} for details)
\begin{equation}\label{P}
\Pi \!=\!N_fT\!\sum_{\omega_n}\!\!\int\!\!
d\xi\left\langle\frac{[i\omega_+-\xi_+] [i\omega_-+\xi_-]+\Delta_h
\Delta_e }{[\xi^2_++\omega^2_++\Delta^2_h]
[\xi^2_-+\omega^2_-+\Delta^2_e ]} \right\rangle_\phi.
\end{equation}
Here $\omega_\pm=\omega_n\pm\Omega_m/2$, $\xi_\pm=\xi\pm\delta/2$,
and we  replaced the integration over momentum $\bm{k}$ by $\int
\ldots d^2\bm{k}/{4\pi^2} = N_f \int \ldots  d \xi d\phi/(2\pi)$,
where $N_f$ is the density of states. Parameter
$\delta=\delta_\phi+\delta_q$ accounts for the doping-induced
modification of the Fermi surfaces. The term
$\delta_\phi=\delta_0+\delta_2\cos2\phi$ describes changes in the
Fermi surfaces radii and overall shape (ellipticity), while the term
$\delta_q=v_Fq\cos(\phi-\psi)$ describes the relative shift in the
centers of Fermi surfaces, where $\phi$ and $\psi$ are the
directions of $\bm{k}_F$ and $\bm{q}$. The magnetic SDW critical
point is determined in terms of doping parameters $\delta_0$ and
$\delta_2$ from the condition $\Gamma=0$, where $\Gamma
=(g^{-1}_{\rm sdw}+\Pi(0,0)) N^{-1}_f$.

We first consider the case of equal gaps on both Fermi surfaces
($\alpha=0$, $\Delta_h = \Delta_e = \Delta$) and then discuss how
the results are modified in the case where the gaps on electron
pockets have nodes. Earlier calculations show~\cite{Vorontsov-SUST}
that there is a broad parameter range
$0.8\lesssim\delta_2/\delta_0\lesssim4.7$ for which SDW order
emerges gradually, and its appearance does not destroy SC order;
i.e., SDW and SC orders coexist over some range of dopings. Since we
are interested in the $T=0$ limit, it is sufficient to evaluate the
propagator of magnetic fluctuations in Eq.~\eqref{L} only at small
frequencies and momenta. A straightforward expansion leads
to~\cite{SM}
\begin{equation}\label{L-approx}
L(\bm{q},\Omega_m)=\frac{1}{N_f}\frac{1}{\eta
v^2_Fq^2+\chi\Omega^2_m+\Gamma},
\end{equation}
where
\begin{subequations}
\begin{equation}\label{Gamma}
\Gamma=\ln\left(\frac{T_{c,0}}{T_{N,0}}\right)-
\left\langle\frac{|\delta_\phi|\arccosh\sqrt{1+\delta^2_\phi/\Delta^2}}
{\sqrt{\delta^2_\phi+\Delta^2}}\right\rangle_\phi,
\end{equation}
\begin{equation}
\chi(\Delta,\delta)=\frac{1}{8}\left\langle\frac{1}{\Delta^2+\delta^2_\phi}
+\frac{\Delta^2\arccosh\left(\sqrt{1+\delta^2_\phi/\Delta^2}\right)}
{|\delta_\phi|(\Delta^2+\delta^2_\phi)^{3/2}}\right\rangle_\phi,
\end{equation}
and
\begin{eqnarray}
\eta(\Delta,\delta,\psi)=\frac{1}{8}\left\langle\cos^2(\phi-\psi)
\left[\frac{2\Delta^2-\delta^2_\phi}{(\Delta^2+\delta^2_\phi)^2}
-3\Delta^2|\delta_\phi|\frac{\arccosh\left(\sqrt{1+\delta^2_\phi/\Delta^2}\right)}
{(\Delta^2+\delta^2_\phi)^{5/2}}\right]\right\rangle_\phi.
\end{eqnarray}
\end{subequations}
In Eq.~(\ref{Gamma}), we absorbed coupling constants $g_{\rm sdw}$
($g_{\rm sc}$) into the corresponding critical temperatures
$T_{N,0}$ ($T_{c,0}$) for the transitions into a pure SDW (SC)
state.

\begin{wrapfigure}{r}{0.45\textwidth}
\begin{center}\vspace{-.55cm}
  \includegraphics[width=0.45\textwidth]{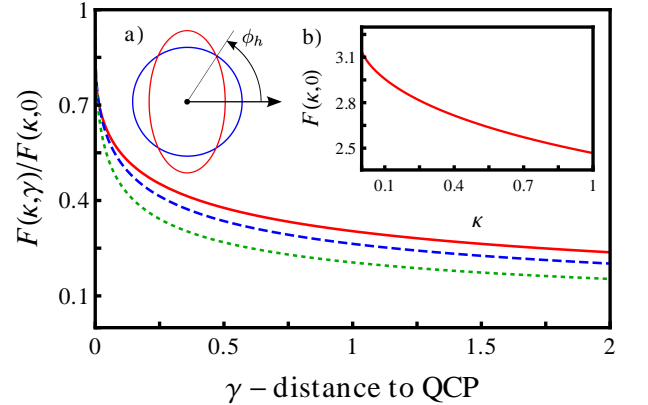}
\end{center}
\caption{Scaling function $F(\kappa,\gamma)$ which accounts for the
interaction correction to the London penetration depth
$\lambda^2/\lambda^2_{BCS} -1 \propto F(\kappa, \gamma)$ is plotted
vs $\gamma$ which measures the distance to the quantum critical
point for three different combinations of the system parameters
encoded by $\kappa=0.1,0.25,0.5$. (see text). Insets: (a) hole
(circular) and electron (elliptical)  Fermi surfaces, $\phi_h$ marks
the location of a hot spot; (b) the dependence of $F(\kappa,0)$,
normalized to the prefactor in Eq.~\ref{ch_99}, on
$\kappa$.}\label{Fig-F}
\end{wrapfigure}

Without superconductivity, $\eta <0$, and a magnetic transition at
$T=0$ is into an incommensurate
phase~\cite{Vorontsov-PRB10,Fernandes-PRB12}. In the presence of SC
order, the commensurate $(\pi,\pi)$ magnetic order is stabilized
($\eta >0$), provided that relevant $\delta^2_\phi \leq \Delta^2$,
which we assume to hold. By order of magnitude, $\chi \sim \eta \sim
1/\Delta^2$.

Substituting Eqs.~\eqref{G} and \eqref{L-approx} into
Eq.~\eqref{Z_1} and integrating explicitly over the momentum
transfer $\bm{q}$ (see Supplementary material for details), we
obtain the fermionic residue for a direction $\phi$ along the Fermi
surface in the form $Z(\phi) =  1 - I(\phi) F$. Here $I(\phi)$
accounts for the (nonsingular) angular dependence and is normalized
such that $(\phi_h)=1$, where $\phi_h$ is the direction of a hot
spot (a $\bm{k}_F$ point for which $\bm{k}_F + \bm{Q}$ is also on
another Fermi surface, see Fig.~\ref{Fig-F}(a), and $F$ accounts for
the dependence on the distance to the hot spot, measured by
$\Gamma$, and on the system parameters $\delta_0$ and $\delta_2$. In
explicit form $F = \langle F(\kappa(\psi),\gamma
(\psi))\rangle_\psi$, where $\kappa(\psi)=\chi/\eta(\psi)$,
$\gamma=\Gamma/\eta(\psi)\Delta^2$, and
\begin{eqnarray}
F(\kappa,\gamma)=\frac{3 }{8\pi^2\eta
N_fv^2_F\Delta}\int^{+\infty}_{-\infty}\frac{\kappa z^2dz}{(1-\kappa)z^2+1-\gamma}\nonumber\\
\times\left[\frac{1}{\kappa
z^2+\gamma}-\frac{\arccosh\left(\sqrt{\frac{z^2+1}{\kappa
z^2+\gamma}}\right)}{\sqrt{z^2+1}\sqrt{(1-\kappa)z^2+1-\gamma}}\right].
\label{ch_99}
\end{eqnarray}
Using Eqs.~(\ref{Z}) and (\ref{Z_1}) we find
\begin{equation}
\beta = \lambda^2/\lambda^2_{BCS} -1 = F \langle I(\phi)\rangle
\end{equation}

Because angular integrals over $\phi$ in $I(\phi)$ and over $\psi$
in $F$ are nonsingular, the dependence of $\beta$ on the distance to
the critical point and on system parameters can be approximated by
$\beta \sim F(\kappa,\gamma)$. It is apparent from the integral in
(\ref{ch_99}) that $F(\kappa,\gamma)$  is finite even in the limit
$\gamma\to0$, which implies that the penetration depth remains
finite at the SDW QCP. Still, $F(\kappa,\gamma)$ is  peaked at the
SDW QCP (when $\gamma=0$), and decreases as $F(\kappa,\gamma)
\propto\ln\gamma/\sqrt{\gamma}$ at $\gamma \gg 1$. We illustrate
this behavior in Fig.~\ref{Fig-F}. Because $\delta \lambda  \sim
F(\kappa,\gamma)$, the penetration depth is also peaked at the QCP.
This behavior is in agreement with the data for isovalent
BaFe$_2$(As$_{1-x}$P$_x$)$_2$~\cite{Matsuda-2}.

By order of magnitude $F(\kappa,0) = O(1)$, hence $\beta = O(1)$.
The enhancement of $\lambda^2 = \lambda^2_{BCS} (1+\beta)$ at the
SDW QCP is larger if magnetic order remains commensurate $(\pi,\pi)$
even in the absence of superconductivity. In this situation,
$\delta_0$ and $\delta_2$ are not restricted to be smaller than
$\Delta$, and, if they are larger, $\beta$ is enhanced by
$(\delta/\Delta)^2$.  We caution, however, that once $\beta$ becomes
large, the one-loop approximation is no longer applicable, and, in
particular, vertex corrections has been analyzed in more
detail~\cite{max_last}.

We next computed $F(\kappa, \gamma)$ for the case when the SC gap
has nodes on electron pockets. We found that, roughly, the angular
dependence of the gap  renormalizes $\kappa$ downward. This,
however, does not change $\lambda$ qualitatively -- at a magnetic
QCP $F(\kappa, 0)$ increases when $\kappa$ decreases, but still
remains finite. We illustrate this in Fig.~\ref{Fig-F}(b). The
reasoning is simple: the nodes of the $s^{+-}$ gap are located at
accidental $\bm{k}_F$ points which generally differ from hot spots.
In the special case where the gap nodes coincide with hot spots, $Z$
at a hot spot diverges logarithmically at a SDW QCP, but the
momentum integral of $Z(\phi)$ is still finite, hence, $\lambda$
remains finite even in this case.

\textit{Conclusions}.-- In this Letter we considered the behavior of
the penetration depth $\lambda (x)$ in a clean Fe-based $s^{+-}$
superconductor at the onset of a commensurate SDW order  inside the
SC phase at $T=0$. We found that the penetration depth remains
finite but has a peak at the onset of SDW order. The magnitude of
the peak is larger when the $s^{+-}$  gap has accidental nodes, but
still remains finite at the onset of SDW order. Our results agree
with the measurements\cite{Matsuda-2} of the penetration depth in
the isovalent BaFe$_2$(As$_{1-x}$P$_x$)$_2$ inside the
superconducting phase. Experiment~\cite{Matsuda-2} shows that that
$\lambda$ has a peak at roughly the same doping where the N\'{e}el
temperature $T_N$ intersects with $T_c$. Our results support the
scenario that SDW order in BaFe$_2$(As$_{1-x}$P$_x$)$_2$ persists
into the SC phase, as happens in other Fe-based superconductors, and
that the peak in the penetration depth occurs at a magnetic
quantum-critical point inside the SC dome. Whether SDW and SC orders
coexist microscopically or phase separate, remains to be seen.

We thank  Y.~Matsuda, T.~Shibauchi, R.~Fernandes, S.~Maiti,
R.~Prozorov, and S.~Sachdev for useful discussions. A.L.
acknowledges support from Michigan State University. M.G.V. is
supported by NSF Grant No. DMR 0955500. A.V.C. is supported by the
DOE Grant No. DE-FG02-ER46900.

\appendix
\section{The model Hamiltonian}

The basic Hamiltonian for electron-electron interaction includes the
free fermion part $H_0$, and four-fermion interaction terms. In the
orbital basis, the free-electron part contains inter-and
inter-orbital hopping terms, and the interaction part  contains
intra-orbital and inter-orbital density-density interactions (the
Hubbard terms $U$ and $U'$), the spin-spin interaction $J$ (the Hund
coupling), and the pair-hopping term $J'$ (Ref.\cite{rev,scal}).

The orbital Hamiltonian is converted into band basis by
diagonalization of the quadratic form. The new quadratic part
describes fermions near hole and electron pockets. We adopt the
minimal three-band model with one hole and two electron pockets. In
this model
 \begin{equation}
H_0=\sum_{\mathbf{k}s}\xi_h(\mathbf{k})c^\dag_{\mathbf{k}s}c_{\mathbf{k}s}
+\sum_{\mathbf{k}'sj}\xi_{je}(\mathbf{k}')f^\dag_{\mathbf{k}'sj}f_{\mathbf{k}'sj}
\end{equation}
where creation/annihilation $c$-operators describe fermions near the
hole pocket, and $f$-operators describe fermions near the electron
pockets labeled by $j=1,2$;   $\xi_h$ and $\xi_{je}$ are
corresponding dispersions. The momenta $\mathbf{k}$ are measured
from the center of the Brilloun zone and $\mathbf{k}'$ are
deviations from $\mathbf{Q} = (\pi,\pi)$.

The four-fermion interaction terms in the band basis are Hubbard,
Hund, and pair-hopping interactions, dressed by coherence factors
from the diagonalization of the quadratic form. There are five
different interaction terms in the band basis~\cite{rev_1} -- two
density-density intra-pocket interactions ($u_4$ and $u_5$ terms in
the notations of~\cite{rev_1}, these interactions are often treated
as equal), density-density inter-pocket interaction $u_1$, exchange
inter-pocket interaction $u_2$, and inter-pocket pair hopping $u_3$.
These five interactions can be re-absorbed into interactions in the
particle-particle channel, and SDW and CDW particle-hole channels.
For repulsive interactions, SDW and SC channels are the two most
relevant ones.  In general, these two interactions may depend on the
directions along the pockets.

For spin-single SC channel, the most relevant interaction is the
pair hopping of two fermions from electron to hole pockets and vice
versa
\begin{equation}
H_\Delta=\frac{1}{2}\sum_{\mathbf{kp}}\sum_{jss'\sigma\sigma'}g^{sc}_{s\sigma\sigma's'}(\mathbf{k},\mathbf{p})
[c^\dag_{\mathbf{k}s}c^\dag_{-\mathbf{k}\sigma}f_{-\mathbf{p}\sigma'j}f_{\mathbf{p}s'j}
+f^\dag_{\mathbf{k}sj}f^\dag_{-\mathbf{k}\sigma
j}c_{-\mathbf{p}\sigma'}c_{\mathbf{p}s}].
\end{equation}
where $g^{sc}_{s\sigma\sigma's'}(\mathbf{k},\mathbf{p})=u_3
(i\tau^y)_{s\sigma}(i\tau^y)^\dag_{\sigma's'}$. Once $u_3$  exceeds
intra-pocket density-density interaction $u_4$, the system develops
a $\pm$ s-wave superconductivity. Since our goal is to analyze the
behavior of the penetration depth deep in the superconducting state,
we do not explicitly solve superconducting problem, but rather
assume that for the doping range we are interested in the system at
$T=0$ is already in the superconducting state.

The magnetic SDW  interaction between fermions
is~\cite{Vorontsov-PRB10}
\begin{equation}
H_\sigma=-\frac{1}{4}\sum_{\mathbf{p}'-\mathbf{p}=\mathbf{k}'-\mathbf{k}}
\sum_{jss'\sigma\sigma'}g^{sdw}_{s\sigma\sigma's'}(\mathbf{pp}',\mathbf{kk}')
[f^\dag_{\mathbf{p}'sj}c_{\mathbf{p}\sigma}c^\dag_{\mathbf{k}\sigma'}f_{\mathbf{k}'s'j}+
f^\dag_{-\mathbf{p}'sj}c_{-\mathbf{p}\sigma}c^\dag_{-\mathbf{k}\sigma'}f_{-\mathbf{k}'s'j}].
\end{equation}
where
$g^{sdw}_{s\sigma\sigma's'}(\mathbf{pp}',\mathbf{kk}')=(u_1+u_3)\bm{\tau}_{s\sigma}
\cdot\bm{\tau}^\dag_{\sigma's'}$, and $\tau$ are the Pauli matrices.

This is the bare  interaction in the SDW channel. The full one is
obtained by dressing this interaction by RPA-type
bubbles~\cite{rev}a.  Such a renormalization (described in detail in
\cite{scal}) transforms a constant interaction  into an effective
interaction mediated by collective spin fluctuations. The static
part of the spin-fluctuation propagator comes from high-energy
fermions, is regular, and can be absorbed into $g_{sdw}$. The
dynamical part of the spin-polarization operator comes from fermions
with low-energies~\cite{acs} and has to be treated exactly. This
last renormalization is given by the sum of bubbles made of normal
and anomalous fermionic Green functions in a superconductor. In
explicit form, the spin-fluctuation propagator is given by Eqs. (5)
and (6) in the main text.

\section{Derivation of $\Gamma$ in Eq.~(8a)}

Distance to the quantum critical point in our model is defined by
$\Gamma = (g^{-1}_{\mathrm{sdw}}-\Pi (0,0))/N_f$, where
$\Pi(\bm{q},\Omega)$ is the polarization operator. Here we provide
the details of the derivation of the equation for $\Gamma$ [Eq.~(8a)
in the main text]. From Eq.~(6) of the main text, we have:
\begin{equation}
\Pi(0,0)=-2\pi T
N_f\sum_{\omega_n>0}\left\langle\frac{E_n}{E^2_n+\delta^2_\phi}\right\rangle_\phi,\qquad
E^2_n=\omega^2_n+\Delta^2.
\end{equation}
where $\langle\ldots\rangle_\phi=\int_0^{2\pi}\ldots d\phi/2\pi$.
Hence
\begin{equation}
\Gamma=\frac{1}{g_{\mathrm{sdw}}N_f}-2\pi T
\sum_{\omega_n>0}\left\langle\frac{E_n}{E^2_n+\delta^2_\phi}\right\rangle_\phi.
\end{equation}
The coupling constant $g_{\mathrm{sdw}}$ can be eliminated in favor
of the transition temperature into the pure SDW state,
$T_{s,0}=(2e^{\gamma_E}/\pi)\Lambda \exp(-1/g_{\mathrm{sdw}}N_f)$,
which allows us to rewrite the equation above as
\begin{equation}
\Gamma=\ln\left(\frac{T}{T_{s,0}}\right)-2\pi T
\sum_{\omega_n>0}\left\langle\frac{E_n}{E^2_n+\delta^2_\phi}-\frac{1}{\omega_n}\right\rangle_\phi.
\end{equation}
Next we add and subtract $1/E_n$ inside the angle brackets and take
the zero temperature limit. The infrared convergent part of the
Matsubara sum can be safely converted into the frequency integral,
while infra-red divergent part is evaluated using  $2\pi
T\sum_{\omega_n}[1/E_n-1/\omega_n]=\ln(\pi e^{-\gamma_E}T/\Delta)$.
This yields
\begin{equation}
\Gamma=\ln\left(\frac{T}{T_{s,0}}\right)+
\left\langle\int^{\infty}_{0}
d\omega\frac{\delta^2_\phi}{\sqrt{\omega^2+\Delta^2}
(\omega^2+\Delta^2+\delta^2_\phi)}\right\rangle_\phi-\ln\left(\frac{\pi
e^{-\gamma_E}T}{\Delta}\right).
\end{equation}
Combining two logarithmic terms and using BCS expression between the
gap and critical temperature $\Delta=\pi e^{-\gamma_E}T_{c,0}$ one
finds for their difference $\ln(T_{c,0}/T_{s,0})$. The remaining
energy integral can be computed analytically with the help of the
formula
\begin{equation}\label{Int}
\int^{\infty}_{0}\frac{dx}{\sqrt{x^2+a^2}(x^2+b^2)}=\frac{\arccosh(b/a)}{|b|\sqrt{b^2-a^2}}.
\end{equation}
Combining last two expressions, we reproduce Eq.~(8a) of the main
text.

\section{Expansion of the polarization operator $\Pi(\bm{q},\Omega_m)$}

To obtain Eq.~(7) of the main text, we need to expand the
polarization operator $\Pi (\bm{q}, \Omega)$ to order $q^2$ and
$\Omega^2$. Expanding the integrand in Eq.~(6) of the main text and
integrating over $\xi$ we obtain
\begin{eqnarray}
&&\Pi(\bm{q},\Omega_m)=\Pi(0,0)+\delta\Pi(0,\Omega_m)+\delta\Pi(\bm{q},0),\\
&&\delta\Pi(0,\Omega_m)=\frac{1}{16}N_fT\Delta^2\Omega^2_m\sum_{\omega_n}\int
d\phi
\frac{\delta^2_\phi+3(\Delta^2+\omega^2_n)}{(\omega^2_n+\Delta^2)^{3/2}
(\omega^2_n+\Delta^2+\delta^2_\phi)^2},\\
&&\delta\Pi(\bm{q},0)=\frac{1}{8} N_fT(v_Fq)^2\sum_{\omega_n}\int
d\phi\frac{\sqrt{\Delta^2+\omega^2_n}(-3\delta^2_\phi+\Delta^2+\omega^2_n)}
{(\omega^2_n+\delta^2_\phi+\Delta^2)^3}.
\end{eqnarray}
At $T \to 0$, Matsubara sums over $\omega_m$ can be converted into
frequency integrals. Integrating over $\omega_m$, we reproduce
Eqs.~(7), (8b) and (8c) of the main text.

\section{Calculation of the self-energy diagram}

To calculate $\lambda/\lambda_{BCS}$, we need to know $ 1 - Z
(\bm{k}_F) = \lim_{\omega\to
0}\partial_{i\omega_m}\Sigma_j(\bm{k}_F,\omega_m)$. The self-energy
is introduced in Eq. (4) of the main text. We have
\begin{equation}
1 - Z (\bm{k}_F) =
3T\lim_{\omega\to0}\sum_{q,\Omega}L(\bm{q},\Omega)
\frac{\partial\mathcal{G}(\bm
{k}_F-\bm{q},\omega-\Omega)}{\partial(i\omega)}
\end{equation}
Using $\partial_{i\omega}=-\partial_{i\Omega}$, taking the limit
$\omega\to0$, and integrating by parts, we obtain
\begin{equation}
1 - Z (\bm{k}_F) =  3T \sum_{q,\Omega}\frac{\partial
L(\bm{q},\Omega)}{\partial(i\Omega)}\mathcal{G}(\bm{k}_F-\bm{q},-\Omega)
\end{equation}
Let's first focus on $\bm{k}_F$ at a hot spot. Using the explicit
forms of the SC Green's function from Eq.~(4) and SDW fluctuation
propagator from Eq.~(7), we express the quasi-particle residue as
\begin{equation}
1-Z(\bm{k}_F)=\frac{6\chi
T}{N_f}\sum_{q,\Omega}\frac{\Omega^2}{(\eta
v^2_Fq^2+\chi\Omega^2+\Gamma)^2(v^2_Fq^2_x+\Omega^2+\Delta^2)}
\label{sm_1}
\end{equation}
The position of $\bm{k}_F$ on the Fermi surface is specified by the
angle $\phi$, $\chi$ and $\Gamma$ are given by Eqs.~(8a) and (8b) in
the main text and depend on system parameters, and $\eta$ is given
by Eqn.~(8c) and depends on system parameters and on the direction
of $\bm{q}$ set by angle $\psi$ (i.e., $\eta = \eta (\psi)$). Taking
the limit of zero temperature and introducing dimensionless
variables
\begin{equation}
x=\frac{v_Fq_x}{\Delta}\quad y=\frac{v_Fq_y}{\Delta}\quad
z=\frac{\Omega}{\Delta}\quad \gamma (\psi) =\frac{\Gamma}{\eta
(\psi)\Delta^2}\quad \kappa (\psi) =\frac{\chi}{\eta(\psi)},
\end{equation}
we re-express Eq.~(\ref{sm_1}) as
\begin{equation}
1-Z(\bm{k}_F)= \left\langle
F(\kappa(\psi),\gamma(\psi))\right\rangle_\psi
\end{equation}
where
\begin{equation}
F(\kappa,\gamma) =
 \frac{3\kappa}{4\pi^3\eta
N_fv^2_F\Delta}\iiint^{+\infty}_{-\infty}\frac{z^2dxdydz}{(x^2+y^2+\kappa
z^2+\gamma)^2(x^2+z^2+1)}.
\end{equation}
Integrating over $y$ and then over $x$, we obtain Eq.~(9) of the
main text.

When $\bm{k}_F$ is not at the hot spot, the expression for the
self-energy becomes more complex, because an intermediate fermion at
$\bm{k}_F + \bm{Q}$ is no longer located at the Fermi surface. This
reduces the self-energy compared to its value at a hot spot.
Roughly, $\Delta$ in Eq.~(\ref{sm_1}) get replaced by
$\sqrt{\Delta^2 + v^2_F k^2_F (\phi)}$, where $k_F(\phi)$ is the
deviation from a hot spot along the corresponding Fermi surface.
With this modification, $1-Z(\bm{k}_F)$ acquires an extra factor
$I(\phi)\approx 1/\sqrt{1 + (k_F(\phi)/k_0)^2}$, where $k_0 =
\Delta/v_F$. At a hot spot, $\phi=\phi_h$ and $I(\phi_h) =1$.
 The parameter $\gamma$ also has to be modified
accordingly, but this is not important for our purposes because we
treat $\gamma$ as a phenomenological parameter which measures the
distance to a magnetic critical point.

It is instructive to analyze in more detail  $F(\kappa,\gamma)$ at
the SDW QCP, when $\gamma=0$. We have
\begin{equation}
F(\kappa,0) = \frac{3}{8\pi^2\eta
N_fv^2_F\Delta}\int^{+\infty}_{-\infty}\frac{\kappa
z^2dz}{(1-\kappa)z^2+1} \left[\frac{1}{\kappa
z^2}-\frac{\mathrm{arccosh}\left(\frac{\sqrt{z^2+1}}{\sqrt{\kappa}
|z|}\right)}{\sqrt{z^2+1}\sqrt{(1-\kappa)z^2+1}}\right].
\end{equation}
One can easily make sure that at $\kappa\ll 1$, the dominant
contribution to the integral comes from small $z\sim\kappa$. For
such $z$, $\arccosh(\sqrt{z^2+1}/\sqrt{\kappa}
|z|)\approx\ln(2/\sqrt{\kappa}|z|)$, and hence
\begin{equation}
F(\kappa,0)\approx\frac{3}{8\pi^2\eta
N_fv^2_F\Delta}\int^{+\infty}_{-\infty}\left[\frac{1}{1+
z^2}-\frac{\kappa
z^2\ln\left(\frac{1}{\sqrt{\kappa}|z|}\right)}{(z^2+1)^2}\right]dz.
\end{equation}
With the logarithmic accuracy this yields
\begin{equation}
F(\kappa,0)\approx\frac{3}{8\pi\eta
N_fv^2_F\Delta}\left[1-\frac{\kappa}{4}
\ln\left(\frac{1}{\kappa}\right)\right],\qquad \kappa\ll1.
\end{equation}
In the opposite limit $\kappa\gg1$ the integral predominantly comes
from $z\sim1/\sqrt{\kappa}$, and we have
\begin{equation}
F(\kappa,0)\approx\frac{1}{12\pi^2\eta
N_fv^2_F\Delta}\frac{\ln\kappa}{\sqrt{\kappa}},\qquad \kappa\gg1.
\end{equation}

\section{Interplay between self-energy and vertex corrections to the current correlation function}

In this section we present calculations to prove the point in the
main text  that  one loop self-energy and vertex corrections to
current correlation function in a superconducting state of an
Fe-pnictide add up rather than cancel. We argue that non-cancelation
is the consequence of the fact that the system lacks Galilean
invariance (electron pockets are located at the corners of the
Brillouin zone), and it holds even if the dispersion within each
band can be approximated by a quadratic one. We contrast
Fe-pnictides to  single-band superconductors with the quadratic
dispersion. In the latter vertex and self-energy corrections  cancel
out, as required by Galilean invariance~\cite{Larkin,Leggett,kim}.
To emphasize this point, we assume that the electronic dispersion
near hole and electron pockets is quadratic and consider two cases
-- only intra-pocket interaction and only inter-pocket interaction.
The first case models Galilean invariant case (hole and electron
pockets do not couple), the second case models $s^{+-}$
superconductivity in Fe-pnictides (hole and electron pockets are
coupled, and the superconducting gaps on the two are determined
self-consistently).

We compute the interaction correction to the function $Q({\vec k})$
introduced in the main text. In the London limit, $Q (\vec{k}
\rightarrow 0) = Q = - c^{-1} \int_{0}^{T^{-1}} d \tau \int d
\vec{x} \langle j_x(\tau) j_x(0) \rangle$, where $\vec{j} =e
\sum_{\vec{p},\alpha} \psi^{\dag}_{\vec{p},\alpha} \vec{v}_{\vec{p}}
\psi_{\vec{p},\alpha}$, and $\alpha =1,2$ accounts for spin
projections. The penetration depth is related to $Q$ as
$\lambda^{-2}=(4\pi/c)Q$.

\begin{figure}[h]
\begin{center}
\includegraphics[scale=0.6]{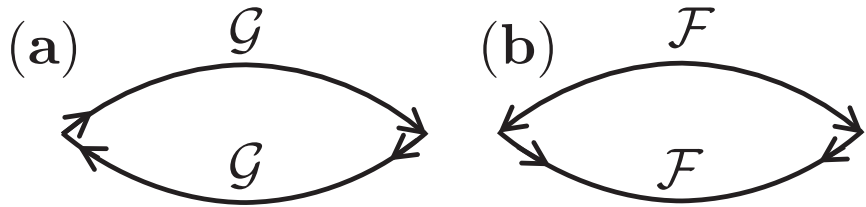}
\caption{Diagrammatic representation of the
 current correlation function in a BCS superconductor.
Diagrams (a) and (b) represent normal and anomalous contributions.}
\label{fig:SM1}
\end{center}
\end{figure}

In the non-interacting case, $Q = Q_0$ is given by the two bubble
diagrams made of the normal and anomalous Green's function,
Fig.~\ref{fig:SM1}. At $T=0$ we have
\begin{align}\label{Q_SM}
Q_0 = \lim_{\vec{k} \to 0} \frac{ 2 e^2 }{c } \frac{ N_F  }{ d }
\int d \xi \int \frac{d \omega}{ 2 \pi} \int \frac{d\theta}{2\pi}
\left\{ v^2_{x,\vec{p}} \left[\mathcal{G} (\xi_+,\omega) \mathcal{G}
(\xi_-,\omega) + \mathcal{F}(\xi_+,\omega) \mathcal{F}(\xi_-,\omega)
- \mathcal{G}_n(\xi_+,\omega) \mathcal{G}_n(\xi_-,\omega) \right]
\right\}\, ,
\end{align}
where $\xi_{\pm} = \xi \pm \vec{v}_{\vec{p}} {\vec k}/2$,  the
overall factor of 2 is due to spin summation, $d$ is the inter-layer
spacing, $N_F = m/(2\pi)$ is the 2D density of states, and
 the normal and anomalous Green's functions $\mathcal{G}$ and $\mathcal{F}$ are
\begin{align}\label{can1_1}
\mathcal{G}(\xi_{\vec{p}},\omega_n) = -\frac{i \omega_n  +
\xi_{\vec{p}}}{\omega_n^2 + \xi_{\vec{p}}^2 + \Delta^2}\, ,\quad
\mathcal{F}(\xi_{\vec{p}},\omega_n) = -\frac{\Delta} {\omega_n^2 +
\xi_{\vec{p}}^2 + \Delta^2 }\, .
\end{align}
We followed Ref. \cite{LP9} and subtracted from $Q_0$ in
(\ref{Q_SM}) its expression in the normal state, (the second term,
with $\mathcal{G}_n (\xi_{\vec{p}}, \omega_n) = 1/(i \omega_n -
\xi_{\vec{p}})$), because there is no superconducting current in the
normal metal. In this form, the integrand is regular and the
integrations over momentum and frequency can be performed in
arbitrary order. We integrate \eqref{Q_SM} over $\xi$ first. The
second term then drops out, and in the first we can safely set
${\vec k} =0$ after the integration. The angular integration yields
$\int (d\theta/2\pi) v^2_{x,\vec{p}} = v^2_F/2 = p^2_F/2m^2$, and
$Q_0$ becomes
\begin{align}\label{Q_SM_A}
Q_0 = \frac{ e^2 }{ m^2 c } \frac{ p_F^2 }{ d } N_F \int_0^\infty d
\omega_n \frac{\Delta^2}{(\omega^2_n + \Delta^2)^{3/2}}\, .
\end{align}
Integrating over $\omega_n$ we obtain
 \begin{align}\label{Q_SM1}
Q_0 = \frac{ e^2 n_s}{ m c }\, ,\,
 \quad n_s =  N_F\frac{p^2_F}{ m d } \, .
\end{align}
Using $N_F = m/(2\pi)$ and the relation between $p_F$ and the full
density $n = p^2_F/(2\pi d)$ we find (at $T=0$) $n_s =n$ and $Q_0 =
ne^2/(mc)$ -- the textbook result. Alternatively, one could
integrate over $\omega_m$ first. Then the first term in
Eq.~\eqref{Q_SM} vanishes at $T=0$, while the second one exactly
reproduces  Eq.~\eqref{Q_SM1} with $n=n_s$ (see e.g.,
Ref.~\cite{Schrieffer}).

\begin{figure}[h]
\begin{center}
\includegraphics[scale=0.6]{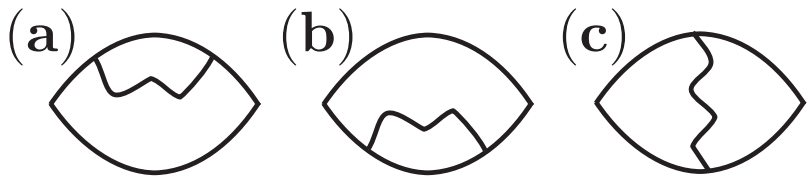}
\caption{Diagrammatic representation of the interaction corrections
to the
 current correlation function.
 Diagrams (a) and (b) represent self energy corrections $\delta Q_{\Sigma}$, and (c) represents vertex correction
$\delta Q_{V}$. The straight lines denote fermion propagators, which
can be either normal or anomalous, and the wavy
 line denotes the boson propagator either in the spin or in the density channel.}
\label{fig:SM0}
\end{center}
\end{figure}

We now proceed with the interacting case. We label interaction
correction to $Q_0$ as $\delta Q$  ($Q = Q_0 + \delta Q$). The total
$\delta Q$ is the sum of self energy and vertex corrections, $\delta
Q = \delta Q_{\Sigma} + \delta Q_{V} $. At one-loop order,  $\delta
Q$ is  represented by the three basic diagrams in
Fig.~\ref{fig:SM0}.

\subsection{Galilean invariant case}

We start with the case when the fermion-fermion interaction is
restricted to the intra-pocket one and consider a single pocket with
a quadratic dispersion. We analyze separately the cases when the
intra-pocket interaction is the density and the spin channels.

To account for the overall factors of different diagrams with normal
and anomalous Green's functions, it is convenient to treat
interaction corrections to $Q$ in the Nambu formalism. We introduce
the Nambu spinor $\chi^{tr}_{\vec{p},\epsilon_n} = [
\chi^{(1)}_{\vec{p},\epsilon_n}, \chi^{(2)}_{\vec{p},\epsilon_n} ] =
[ \psi_{\uparrow; \vec{p}, \epsilon_n}, \psi^{\dag}_{\downarrow;
-\vec{p}, -\epsilon_n}]$. We denote by $\tau_x,\tau_y,\tau_z$ the
Pauli matrices operating in Nambu space along with the unit matrix
$\tau_0 = 1$.
 The current operator in Nambu representation is
$\vec{j} =e \sum_{\vec{p}} \chi^{\dag}_{\vec{p}} \vec{v}_{\vec{p}}
\tau_0 \chi_{\vec{p}}$.
 The non-interacting Nambu Green function is
\begin{align}\label{can1}
\hat{G}(\vec{p},\omega_n) = \frac{ i \omega_n \tau_0 + \xi_{\vec{p}}
\tau_z + \Delta \tau_x } {\omega_n^2 + \xi_{\vec{p}}^2 + \Delta^2
}\, .
\end{align}
The Nambu Green function \eqref{can1} is related to the normal and
anomalous Green functions as
\begin{align}\label{GN}
\hat{G}(\vec{p},\omega_n) =
\begin{bmatrix}
\mathcal{G}(\vec{p},\omega_n) & - \mathcal{F}(\vec{p},\omega_n) \\
- \mathcal{F}(\vec{p},\omega_n) & -\mathcal{G}(-\vec{p},-\omega_n)
\end{bmatrix}\, .
\end{align}

\subsubsection{Interaction in the density channel}

Consider first the case when the dominant intra-pocket interaction
is between fermion densities. Such an interaction is generally
described by the effective action $S_{int} = -(1/2)\sum_{\vec{q},m}
L^{\rho}_{\vec{q},\Omega_m} \rho_{\vec{q},\Omega_m}
\rho_{-\vec{q},-\Omega_m} $, where $\rho (\vec{q}, \Omega_m) =
T\sum_{\vec{p},\omega_m,\alpha}
\psi^{\dagger}_{\vec{p},\omega_m,\alpha}
\psi_{\vec{p}+\vec{q},\omega_+\Omega_m,\alpha}$ ( $\alpha =1,2$
accounts for spin). In the Nambu representation,  $\rho_{\vec{q}} =
\sum_{\vec{p}} \chi^{\dag}_{\vec{p}+\vec{q}} \tau_z \chi_{\vec{p}}$.
The function  $L^{\rho}(\vec{q},\Omega_m)$ is an effective
propagator of charge fluctuation which in the case of intra-pocket
interaction is peaked at $\vec{q}=0,\Omega_m=0$ and decreases when
each of these two parameters increases.

\begin{figure}[h]
\begin{center}
\includegraphics[scale=0.6]{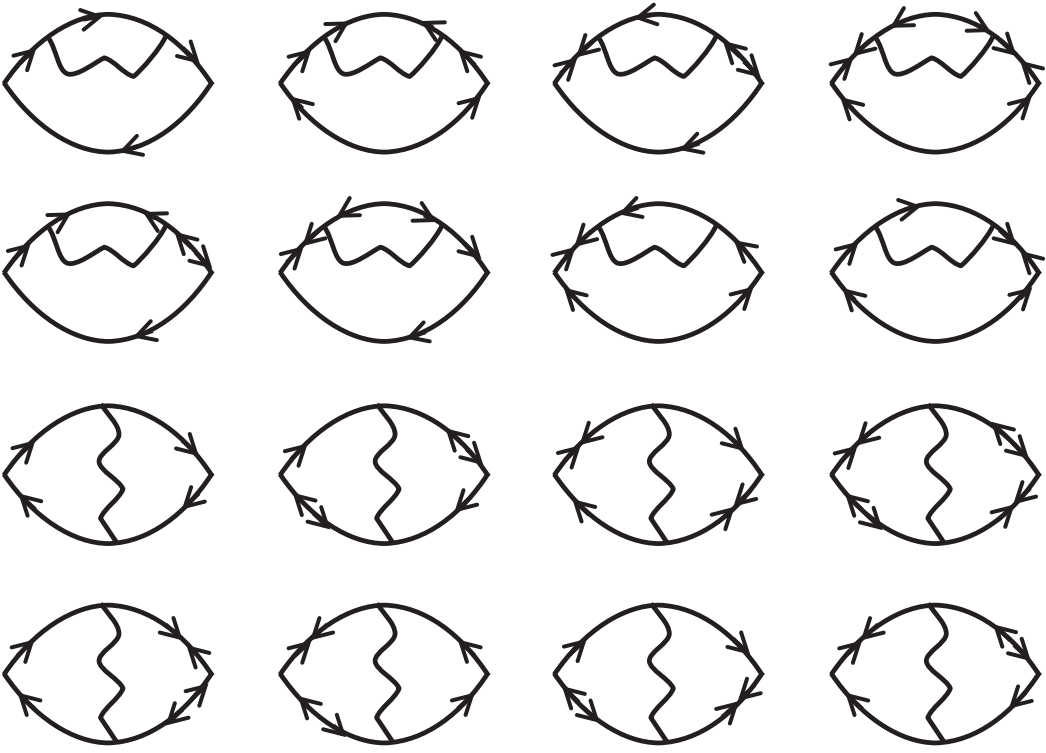}
\caption{Diagrammatic representation of the self energy,
Eq.~\eqref{can2}, (first and second row) and vertex corrections,
Eqs.~\eqref{can5}, \eqref{can6} (third and fourth rows). Fermion
propagators with a single arrow stand for a normal Green functions,
$\mathcal{G}$. The double arrowed lines designate the anomalous
Green functions, $\mathcal{F}$. These contributions  (with the
proper signs) are obtained after taking the trace in the expressions
for $\Delta_Q$ in the Nambu formalism, Eqs.~\eqref{can2} and
\eqref{can6}. } \label{fig:16}
\end{center}
\end{figure}

The self energy correction to the static current correlation
function is
\begin{align}\label{can2}
\delta Q_{\Sigma} = 2  e^2 c^{-1}d^{-1} T \sum_{\vec{p},\omega_n} T
\sum_{\vec{q}, \Omega_m} L^\rho(\vec{q},\Omega_m) \mathrm{Tr}
\left[ v_{x,\vec{p}} \hat{G}(\vec{p},\omega_n) \tau_z
\hat{G}(\vec{p}+\vec{q},\omega_n+\Omega_m) \tau_z
\hat{G}(\vec{p},\omega_n) v_{x,\vec{p}} \hat{G}(\vec{p},\omega_n)
\right]\, ,
\end{align}
where $\mathrm{Tr}$ stands for trace operation in Nambu space, and
the factor of $2$ accounts for two self-energy diagrams in
Fig.~\ref{fig:SM0}. Once the trace over the Nambu indices is taken,
$\delta Q_{\Sigma}$ can be represented diagrammatically by graphs
shown in the first two lines in Fig.~\ref{fig:16}.
Because our primary goal is  to understand when vertex and
self-energy corrections cancel out and when they add up, we make an
additional simplifying assumption that the characteristic momenta
and frequency, above which charge propagator
$L^{\rho}(\vec{q},\Omega_m)$ decreases, are much smaller than
$\Delta/v_{F}$ and $\Delta$, respectively. In this situation,  the
integrations over the fermion and boson energies and momenta
separate, and Eq.~\eqref{can2} takes the form
\begin{align}\label{can3}
\delta Q_{\Sigma}  = e^2 c^{-1} d^{-1}  L_0^{\rho} I_{\Sigma}\, ,
\end{align}
where
\begin{align}\label{can3a}
L_0^{\rho} = T \sum_{\vec{q},\Omega_m}  L^{\rho}(\vec{q},\Omega_m) =
L^{\rho}(\vec{x}=0,t =0)
\end{align}
is an instantaneous interaction, and
\begin{align}\label{can4}
I_{\Sigma} = 2   T \sum_{\vec{p},\omega_n} \mathrm{Tr} \left[
v_{\vec{p},x} \hat{G}(\vec{p},\omega_n) \tau_z
\hat{G}(\vec{p},\omega_n) \tau_z \hat{G}(\vec{p},\omega_n)
v_{\vec{p},x} \hat{G}(\vec{p},\omega_n) \right]\, .
\end{align}
Taking the trace in Eq.~\eqref{can4}  with Eq.~\eqref{GN} we obtain
explicitly,
\begin{align}\label{can4a}
I_{\Sigma} = \langle \vec{v}^2_{\vec{p}}\rangle T
\sum_{\vec{p},\omega_n} \left[\mathcal{G}^4 +\mathcal{G}_-^4 - 2
\mathcal{F}^4  \right]\, ,
\end{align}
where $\mathcal{G}_-(\vec{p},\omega_n) =
\mathcal{G}(-\vec{p},-\omega_n)$ and frequency and momenta arguments
are omitted for clarity.

For the vertex correction part we obtain, within the same
approximation
\begin{align}\label{can5}
\delta Q_{V} =e^2 c^{-1} d^{-1} L_0^{\rho} I_{V}\, ,
\end{align}
where
\begin{align}\label{can6}
I_{V} =  T \sum_{\vec{p},\omega_n} \mathrm{Tr} \left[
\hat{G}(\vec{p},\omega_n) v_{\vec{p},x} \hat{G}(\vec{p},\omega_n)
\tau_z \hat{G}(\vec{p},\omega_n)  v_{\vec{p},x}
\hat{G}(\vec{p},\omega_n) \tau_z \right].
\end{align}
Taking again the trace in Nambu space, we obtain the diagrams
presented in the last two lines in Fig. \ref{fig:16}. Evaluating
them, we find
\begin{align}\label{can6a}
I_{V} =   \frac{\langle \vec{v}^2_{\vec{p}}\rangle }{ 2 } T
\sum_{\vec{p},\omega_n} \left[ \mathcal{G}^4 +\mathcal{G}_-^4 + 2
\mathcal{F}^4 + 4 \mathcal{F}^2 \mathcal{G} \mathcal{G}_- \right]\,
.
\end{align}
At $T=0$ the frequency summation is replaced by integration.
Evaluating frequency and momentum integrals by, e.g., by using polar
coordinates, we obtain
\begin{align}\label{integs}
T \sum_{\vec{p},\omega_n}  \mathcal{F}^4 = 2 T
\sum_{\vec{p},\omega_n} \mathcal{F}^2 \mathcal{G} \mathcal{G}_- =
\frac{  N_F }{ 6 \Delta^2}\, , \quad \,\,\, T
\sum_{\vec{p},\omega_n} \mathcal{G}^4  = T \sum_{\vec{p},\omega_n}
\mathcal{G}_-^4  = T \sum_{\vec{p},\omega_n} \mathcal{F}^2
\mathcal{G}^2 = T \sum_{\vec{p},\omega_n} \mathcal{F}^2
\mathcal{G}_-^2 = 0\, .
\end{align}
This gives
\begin{align}\label{can7}
I_{\Sigma} = - I_{V} = \frac{1}{3} \frac{ \langle
\vec{v}_{\vec{p}}^2 \rangle N_F }{ \Delta^2 }\, .
\end{align}
We see that self-energy and vertex corrections to the static current
correlation function in an s-wave superconductor cancel each other,
as it indeed should be the case because of Galilean invariance.

\subsubsection{Interaction in the spin channel}

We now check that the cancelation in the Galilean invariant case
holds also for the case when the intra-pocket interaction in the
spin channel. The effective action is now $S_{int} =-
(1/2)\sum_{\vec{q},m} L^{\sigma}_{\vec{q},\Omega_m}
\vec{\sigma}_{\vec{q},\Omega_m} \vec{\sigma}_{-\vec{q},-\Omega_m} $.
The Nambu formalism is inconvenient in this case as the spin
operator doesn't have a simple form of a product of a Nambu field
operator and its conjugate. This formal difficulty can be alleviated
by introducing Balian-Werthammer (BW)~\cite{BW} extension of the
Nambu formalism.  The four-dimensional BW spinors are defined as
\begin{align}\label{BW1}
\Psi_{\vec{p}} = \left[ \Psi_{11},\Psi_{12}, \Psi_{21},\Psi_{22}
\right]^{tr} = \left[  c_{\vec{p},\uparrow}, c_{\vec{p},\downarrow},
-c^{\dag}_{-\vec{p},\downarrow}, c^{\dag}_{-\vec{p},\uparrow}
\right]^{tr}\, .
\end{align}
These spinors
 form a tensor product of Nambu space and original spin space.
Let $\sigma_x, \sigma_y, \sigma_z$ be spin Pauli matrices, and
$\sigma_0$  the unit matrix.
 The current, density, and spin density operators are
 $\rho = (1/2)\sum_{\vec{p}} \Psi_{\vec{p}}^{\dag} \check{\rho} \Psi_{\vec{p}}$,
$\vec{j} = e \vec{v}$, $\vec{v}  = (1/2)\sum_{\vec{p}}
\Psi_{\vec{p}}^{\dag} \check{\vec{V}}_{\vec{p}} \Psi_{\vec{p}}$ and
$\bm{\sigma}  = (1/2)\sum_{\vec{p}} \Psi_{\vec{p}}^{\dag}
\check{\bm{\sigma}} \Psi_{\vec{p}}$, where
 \begin{align}\label{BW2}
\check{\rho}_{\vec{p}} = \tau_z \otimes \sigma_0\, ,\,\,\,
\check{\vec{V}}_{\vec{p}} = \vec{v}_{\vec{p}} \tau_0 \otimes
\bm{\sigma}_0\, , \,\,\, \check{\bm{\sigma}} =  \tau_0 \otimes
\bm{\sigma}\, ,
\end{align}
and
 $\otimes$ stands for a tensor product. The Green
function in BW space is
 expressed via the Nambu Green
function, Eq. \eqref{can1}, as
\begin{align}\label{BW3}
\check{G} = \hat{G}\otimes \sigma_0\, .
\end{align}
The main advantage of the BW formalism is that it allows one to use
standard diagrammatic rules, including $(-1)$ rule for each closed
loop.

We found that Eqs~\eqref{can3}, \eqref{can3a}, and \eqref{can5}
preserve their form, with  $L^{\sigma}$ instead of $L^{\rho}$, but
Eqs.~\eqref{can4} and \eqref{can6} become, respectively
\begin{align}\label{BW4}
I_{\Sigma} =  T \sum_{\vec{p},\omega_n} \mathrm{Tr} \left[
\check{V}_{\vec{p},x}  \check{G}(\vec{p},\omega_n)
\check{\bm{\sigma}} \check{G}(\vec{p},\omega_n) \check{\bm{\sigma}}
\check{G}(\vec{p},\omega_n) \check{V}_{\vec{p},x}
\check{G}(\vec{p},\omega_n) \right]
\end{align}
and
\begin{align}\label{BW5}
I_{V}=\frac{1}{2} T \sum_{\vec{p},\omega_n} \mathrm{Tr} \left[
\check{G}(\vec{p},\omega_n) \check{V}_{\vec{p},x}
\check{G}(\vec{p},\omega_n) \check{\bm{\sigma}}
 \check{G}(\vec{p},\omega_n)\check{V}_{\vec{p},x}  \check{G}(\vec{p},\omega_n) \check{\bm{\sigma}} \right].
\end{align}
An extra factor of $1/2$ as compared to \eqref{can4} and
\eqref{can6} compensates for
 the double counting of degrees of freedom in BW formalism.
  Taking the trace in
 \eqref{BW4} and \eqref{BW5}
  we obtain
\begin{align}\label{BW4a}
I_{\Sigma} = 3  \langle \vec{v}^2_{\vec{p}}\rangle T
\sum_{\vec{p},\omega_n}
 \left[ 2 \mathcal{F}^4
 +4 \mathcal{F}^2 \left(\mathcal{G}^2-\mathcal{G}
   \mathcal{G}_-  +\mathcal{G}_{-}^2\right)+\mathcal{G}^4+\mathcal{G}_{-}^4
   \right]\, ,
\end{align}
\begin{align}\label{BW5a}
I_V = 3  \frac{\langle \vec{v}^2_{\vec{p}}\rangle }{ 2 } T
\sum_{\vec{p},\omega_n}
 \left[ 2 \mathcal{F}^4
 +4 \mathcal{F}^2 \left(\mathcal{G}^2-\mathcal{G}
   \mathcal{G}_-  +\mathcal{G}_{-}^2\right)+\mathcal{G}^4+\mathcal{G}_{-}^4
   \right]
\end{align}
respectively.
  Using Eq.~\eqref{integs}  we immediately find that
\begin{align}
I_{\Sigma} = I_{V} = 0,
\end{align}
i.e., self energy and vertex corrections simply vanish.

\subsection{The case of inter-pocket interaction}

The basic formalism set up in the previous subsection can be applied
to  the case when the interactions
$L^{\rho,\sigma}(\vec{q},\Omega_m)$  are peaked at large momentum
$\vec{q} =\vec{Q}$, and couple hole and electron pockets. We
consider the simplest situation of perfect nesting, when the
electron and the hole dispersions are  $\xi^e_{\vec{p}} = -
\xi^h_{\vec{p}+\vec{Q}} = \xi_{\vec{p}}$. In this case, the velocity
of a fermions at a point ${\bf k}$ on the hole pocket is opposite to
that of a fermion at ${\bf k} + {\bf Q}$ on the electron pocket. We
also assume that $\Delta_e = - \Delta_h = \Delta$. Finally, like in
the previous case, we assume that
$L^{\rho,\sigma}(\vec{q},\Omega_m)$ are rapidly decaying functions
of $\vec{q}-\vec{Q}$ and $\Omega_m$, and that characteristic
$\vec{q}-\vec{Q}$ and $\Omega_m$ much smaller than $\Delta/v_F$ and
$\Delta$, respectively. This again allows us to factorize  the
integrations over fermion energies and momenta and over boson energy
and momentum.

\subsubsection{Interaction in the density channel}

The electron and hole Nambu Green functions are
\begin{align}\label{Ncan1}
\hat{G}^{e,h}(\vec{p},\omega_n) = \frac{ i \omega_n \tau_0 \pm
\xi_{\vec{p}} \tau_z \pm \Delta \tau_x } {\omega_n^2 +
\xi_{\vec{p}}^2 + \Delta^2 }\, .
\end{align}
Equation \eqref{can3} should
 be modified to include both hole and electron pockets and becomes
\begin{align}\label{Ncan3}
\delta Q_{\Sigma} = \frac{e^2 }{ d c} L_{\pi}^{\rho} ( I^e_{\Sigma}
+ I^h_{\Sigma} )\, ,
\end{align}
where
\begin{align}\label{can3a1}
L_{\pi}^{\rho} = T \sum_{\vec{q} \approx {\vec Q},\Omega_m}
L^{\rho}(\vec{q},\Omega_m)
\end{align}
and
\begin{align}\label{Ncan4}
I^{e(h)}_{\Sigma} = 2   T \sum_{\vec{p},\omega_n} \mathrm{Tr} \left[
v^{e(h)}_{\vec{p},x} \hat{G}^{e(h)}(\vec{p},\omega_n) \tau_z
\hat{G}^{h(e)}(\vec{p},\omega_n) \tau_z
\hat{G}^{e(h)}(\vec{p},\omega_n) v^{e(h)}_{\vec{p},x}
\hat{G}^{e(h)}(\vec{p},\omega_n) \right].
\end{align}
Taking the trace and counting pre-factors, we obtain
\begin{align}\label{Ncan4a}
I^{e}_{\Sigma} =I^{h}_{\Sigma} =  \langle \vec{v}_{\vec{p}}^2
\rangle
 T \sum_{\vec{p},\omega_n}
\Big[ - &2 \mathcal{F}_{e}^3 \mathcal{F}_{h}+\mathcal{F}_{e}^2 (2
\mathcal{G}_{e} \mathcal{G}_{h}-\mathcal{G}_{e}
   \mathcal{G}_{h,-}-\mathcal{G}_{e,-} \mathcal{G}_{h}+2
   \mathcal{G}_{e,-} \mathcal{G}_{h,-})
\notag \\
   -& 2 \mathcal{F}_{e}
   \mathcal{F}_{h} \left(\mathcal{G}_{e}^2-\mathcal{G}_{e}
  \mathcal{G}_{e,-}+\mathcal{G}_{e,-}^2\right)+\mathcal{G}_{e}^3
   \mathcal{G}_{h}+\mathcal{G}_{e,-}^3 \mathcal{G}_{h,-}
   \Big]\, .
\end{align}
The non-zero contributions are
\begin{align}\label{Ncan4b}
T \sum_{\vec{p},\omega_n} \mathcal{F}_{e}^3 \mathcal{F}_{h} = 2 T
\sum_{\vec{p},\omega_n} \mathcal{F}_{e}^2 \mathcal{G}_{e}
\mathcal{G}_{h} = 2 T \sum_{\vec{p},\omega_n} \mathcal{F}_{e}^2
\mathcal{G}_{e,-} \mathcal{G}_{h,-} = 2 T \sum_{\vec{p},\omega_n}
\mathcal{F}_{e} \mathcal{F}_{h} \mathcal{G}_{e} \mathcal{G}_{e,-} =
- \frac{ N_F }{ 6 \Delta^2}\, .
\end{align}
Substituting this into (\ref{Ncan4a}), we obtain
\begin{align}\label{Ncan7}
I^e_{\Sigma} = I^h_{\Sigma} = I^{eh}_{V} = I^{he}_{V} = -\frac{1}{6}
\frac{ \langle \vec{v}_{\vec{p}}^2 \rangle N_F }{ \Delta^2 }.
\end{align}

The vertex correction contribution so the current-current
correlation function is
\begin{align}\label{Ncan5}
\delta Q_{V}= \frac{e^2}{c d} L_{\pi}^{\rho} (I^{eh}_{V} +
I^{he}_{V})\, ,
\end{align}
where now
 \begin{align}\label{Ncan6}
I^{eh(he)}_{V} =  T \sum_{\vec{p},\omega_n} \mathrm{Tr} \left[
\hat{G}^{e(h)}(\vec{p},\omega_n) v^{e(h)}_{\vec{p},x}
\hat{G}^{e(h)}(\vec{p},\omega_n) \tau_z
 \hat{G}^{h(e)}(\vec{p},\omega_n)  v^{h(e)}_{\vec{p},x}  \hat{G}^{h(e)}(\vec{p},\omega_n) \tau_z \right].
\end{align}
 After taking the trace, this becomes
 \begin{align}\label{Ncan6a}
I^{eh}_{V} = I^{he}_{V} = - \frac{ \langle \vec{v}_{\vec{p}}^2
\rangle  }{ 2 }T \sum_{\vec{p},\omega_n} \Big[
 & \mathcal{G}_e^2 \mathcal{G}_h^2
 + \mathcal{G}_{e,-}^2 \mathcal{G}_{h,-}^2
 + 2 \mathcal{F}_e^2 \mathcal{F}_h^2 +
\mathcal{F}_h^2 ( \mathcal{G}_{e}^2 + \mathcal{G}_{e,-}^2) +
 \mathcal{F}_e^2 ( \mathcal{G}_h^2 + \mathcal{G}_{h,-}^2)
\notag \\
& - 2 \mathcal{F}_h \mathcal{F}_e \left(
 \mathcal{G}_{e} \mathcal{G}_{h}
 +  \mathcal{G}_{e,-} \mathcal{G}_{h,-}
-   \mathcal{G}_{e,-} \mathcal{G}_{h} - \mathcal{G}_{e}
\mathcal{G}_{h,-} \right)\Big].
\end{align}
The non-zero terms in \eqref{Ncan6a} are
 \begin{align}\label{Ncan6b}
T \sum_{\vec{p},\omega_n} \mathcal{G}_e^2 \mathcal{G}_h^2 = T
\sum_{\vec{p},\omega_n} \mathcal{G}_{e,-}^2 \mathcal{G}_{h,-}^2 = T
\sum_{\vec{p},\omega_n} \mathcal{F}_e^2 \mathcal{F}_h^2 = 2 T
\sum_{\vec{p},\omega_n} \mathcal{F}_h \mathcal{F}_e \mathcal{G}_{e}
\mathcal{G}_{h} = 2 T \sum_{\vec{p},\omega_n} \mathcal{F}_h
\mathcal{F}_e \mathcal{G}_{e,-} \mathcal{G}_{h,-} = \frac{ N_F }{ 6
\Delta^2 }\, .
\end{align}
 Substituting into \eqref{Ncan6a}, we obtain
\begin{align}\label{Ncan7_1}
 I^{eh}_{V} = I^{he}_{V} = -\frac{1}{6} \frac{ \langle \vec{v}_{\vec{p}}^2 \rangle N_F }{ \Delta^2 }\, .
\end{align}
Evidently, $\delta Q_{\Sigma} = \delta Q_{V}$, i.e., self-energy and
vertex corrections  add up rather than cancel out. By obvious
reasons, non-cancelation survives even if we relax the assumption of
a perfect nesting and of exactly opposite values of the gaps on hole
and electron pockets, and do not require that
$L^{\rho,\sigma}(\vec{q},\Omega_m)$ are rapidly decaying functions
of $\vec{q}-\vec{Q}$ and $\Omega_m$.

The total $\delta Q = \delta Q_{\Sigma} + \delta Q_{V}$ is
\begin{align}\label{Ncan8}
\delta Q = -\frac{2}{3} \frac{ e^2}{d c} L_{\pi}^{\rho} \frac{
\langle \vec{v}_{\vec{p}}^2 \rangle N_F }{ \Delta^2 }\, .
\end{align}
 The relative correction to the penetration depth
\begin{align}\label{rel}
\frac{ \delta \lambda }{ \lambda_0 } =-\frac{\delta Q}{ 2 Q_0 } =
\frac{L_{\pi}^{\rho} }{6 \Delta^2 }\, ,
\end{align}
where $Q_0$ is given by Eq.~\eqref{Q_SM1}, and the factor of 2 is
because we include both electron and hole pockets. We see that the
corrections tend to increase the penetration depth. The
instantaneous interaction $L_{\pi}^{\rho}$, given by
Eq.~\eqref{can3a}, is of order $H_{int} t_{ret}^{-1}$, where
$H_{int}$ is the typical local interaction strength, and $t_{ret}$
is the retardation time. Equation \eqref{rel} shows that the
perturbation theory is valid provided, $H_{int} \ll \Delta (t_{ret}
\Delta)$. Because we assumed $t_{ret} \gg \Delta^{-1}$, the
perturbative approach is valid even if the interaction exceeds
$\Delta$.

\subsubsection{Interaction in the spin channel}

Consider next inter-pocket interaction in the spin channel.
 Applying the BW formalism,
 we obtain, after straightforward calculations
\begin{align}\label{Ncan9}
I^{e(h)}_{\Sigma} = T \sum_{\vec{p},\omega_n} \mathrm{Tr} \left[
\check{V}^{e(h)}_{\vec{p},x} \check{G}^{e(h)}(\vec{p},\omega_n)
\check{\bm{\sigma}} \check{G}^{h(e)}(\vec{p},\omega_n)
\check{\bm{\sigma}} \check{G}^{e(h)}(\vec{p},\omega_n)
\check{V}^{e(h)}_{\vec{p},x}  \check{G}^{e(h)}(\vec{p},\omega_n)
\right]\, ,
\end{align}
and
\begin{align}\label{Ncan10}
I^{eh(he)}_{V} =  \frac{1}{2} T \sum_{\vec{p},\omega_n} \mathrm{Tr}
\left[ \check{G}^{e(h)}(\vec{p},\omega_n)
\check{V}^{e(h)}_{\vec{p},x} \check{G}^{e(h)}(\vec{p},\omega_n)
\check{\bm{\sigma}}
 \check{G}^{h(e)}(\vec{p},\omega_n) \check{V}^{(h)e}_{\vec{p},x}
 \check{G}^{(h)e}(\vec{p},\omega_n) \check{\bm{\sigma}}
  \right]\, .
\end{align}
 Taking the trace over spin and Nambu indices we get
\begin{align}\label{Ncan9a}
I_{\Sigma}^e = I_{\Sigma}^h =  3 \langle \vec{v}_{\vec{p}}^2 \rangle
T \sum_{\vec{p},\omega_n} \Big[ 2 \mathcal{F}_e^3 \mathcal{F}_h &+ 2
\mathcal{F}_e \mathcal{F}_h (\mathcal{G}_e^2 - \mathcal{G}_e
\mathcal{G}_{e,-}+ \mathcal{G}_{e,-}^2) + \mathcal{G}_e^3
\mathcal{G}_h + \mathcal{G}_{e,-}^3 \mathcal{G}_{h,-}
\notag \\
& + \mathcal{F}_e^2 (2 \mathcal{G}_e \mathcal{G}_h -
\mathcal{G}_{e,-} \mathcal{G}_h - \mathcal{G}_e \mathcal{G}_{h,-} +
2 \mathcal{G}_{e,-} \mathcal{G}_{h,-})
    \Big]\, ,
\end{align}
\begin{align}\label{Ncan10a}
I_{V}^{eh} = I_{V}^{he} =  -3 \frac{ \langle \vec{v}_{\vec{p}}^2
\rangle}{2} T \sum_{\vec{p},\omega_n} \Big[ \mathcal{F}_h^2
(\mathcal{G}_e^2 + \mathcal{G}_{e,-} ^2) &+ \mathcal{G}_e^2
\mathcal{G}_h^2 + 2 \mathcal{F}_e \mathcal{F}_h (\mathcal{G}_e
\mathcal{G}_h + \mathcal{G}_{e,-}  \mathcal{G}_{h,-} -
\mathcal{G}_{e,-} \mathcal{G}_h - \mathcal{G}_e \mathcal{G}_{h,-} )
\notag \\
& +
   \mathcal{G}_{e,-} ^2 \mathcal{G}_{h,-}^2 +
   \mathcal{F}_e^2 (2 \mathcal{F}_h^2 + \mathcal{G}_h^2 + \mathcal{G}_{h,-}^2)
\Big]
\end{align}
The non-zero terms in
 Eqs.~\eqref{Ncan9a} and \eqref{Ncan10a} are listed in Eqs.~\eqref{Ncan4b} and \eqref{Ncan6b}.
 Substituting them, we obtain
\begin{align}\label{Ncan11}
I^{e}_{\Sigma} = I^{h}_{\Sigma} = I^{eh}_{V}= I^{he}_{V} =- 3 \frac{
\langle \vec{v}_{\vec{p}}^2 \rangle N_F }{2 \Delta^2}\, .
\end{align}
We see that self-energy and vertex corrections again add up.
Combining the two, we obtain
\begin{align}\label{Ncan12}
\delta Q = -6\frac{e^2}{d c} L_{\pi}^{\sigma} \frac{  \langle
\vec{v}_{\vec{p}}^2 \rangle N_F }{ \Delta^2 }\, .
\end{align}
 The correction to the penetration
 depth
 is
\begin{align}\label{rel1}
\frac{ \delta \lambda }{ \lambda_0 } =-\frac{\delta Q}{ 2 Q_0 } =
\frac{3 L_{\pi}^{\sigma} }{2 \Delta^2 }\, .
\end{align}

\subsubsection{Aslamazov-Larkin diagrams}

In the calculations of the  current correlation function in the
normal state, it is important to include into consideration also
Aslamazov-Larkin diagrams~\cite{Larkin-Varlamov-Book}, which contain
two boson propagators $L$ coupled via two triangular fermionic
vertex blocks. We analyzed these diagrams for our case and found
that they are irrelevant. First, for the case when the gaps on hole
and electron FSs are of equal magnitude (and opposite sign), there
is a cancelation of the dominant Aslamazov-Larkin terms. The
cancelation occurs at the level of the valuation of the fermionic
triangular vertex -- for each block there are two ways to arrange
electron and hole Green's function lines and their corresponding
momenta which cancel each other. For non-equal gap magnitudes, the
cancelation is not prefect. Still, even in this case, the
Aslamazov-Larkin terms can be safely neglected. The reason is the
following: the overall factor in Aslamazov-Larkin terms contains two
additional powers of the coupling constant. In the normal state,
these two additional powers and eliminated by the need to regularize
the divergence coming out of integration of the two bosonic
propagators. In the superconducting state,  the corresponding
integral is finite because for small $q$ each triangular block
scales as $v_F q/\Delta$.  As a result, the two extra powers of the
coupling do not cancel.

\end{document}